\documentclass[letterpaper,aps,twocolumn,floats,showpacs,preprintnumbers,nofootinbib,prl]{revtex4}
\usepackage{amsmath,amssymb}
\usepackage{epsfig}
\usepackage{bbm}
\usepackage[dvips]{hyperref}

\newcommand{\D}{{\rm d}}
\begin{document}

\title{The Hubble diagram as a probe of mini-charged particles}
\author{Markus Ahlers}
\affiliation{Rudolf Peierls Centre for Theoretical Physics, University of Oxford, Oxford OX1 3NP, UK}

\preprint{OUTP-0908P}

\begin{abstract}
The luminosity--redshift relation of cosmological standard candles provides information about the relative energy composition of our Universe. In particular, the observation of type Ia supernovae up to redshift of $z\sim2$ indicate a universe which is dominated today by dark matter and dark energy. The propagation distance of light from these sources is of the order of the Hubble radius and serves as a very sensitive probe of feeble inelastic photon interactions with background matter, radiation or magnetic fields. In this paper we discuss the limits on mini-charged particle models arising from a dimming effect in supernova surveys. We briefly speculate about a strong dimming effect as an alternative to dark energy.
\end{abstract}


\maketitle

\section{Motivation}

In recent years the increasing amount and accuracy of astronomical data has improved our knowledge about the composition and evolution of the Universe dramatically. Measurements of the temperature anisotropies of the cosmic microwave background, supernova surveys and the analysis of the power spectrum of galaxy clustering are in concordance with a spatially flat universe, which has recently become dominated by vacuum energy $\Lambda$ and cold dark matter (CDM) -- the so-called $\Lambda$CDM model (for a review see Ref.~\cite{Amsler:2008zzb}). With the increasing precision of cosmological parameters it is feasible that exotic particle interactions with a tiny rate $\Gamma$ comparable to the Hubble expansion rate $H$ can be tested in cosmological surveys.

In particular, luminosity distance measurements of cosmological standard candles like type Ia supernovae (SNe)~\cite{Riess:1998cb,Kowalski:2008ez} might test feeble photon interactions with background magnetic fields, radiation and matter. The luminosity distance $d_L$ is defined as 
\begin{equation}\label{dL}
d_L(z) \equiv \sqrt{\frac{\mathcal{L}}{4\pi F}}\,,
\end{equation}
where $\mathcal{L}$ is the luminosity of the standard candle (assumed to be sufficiently well-known) and $F$ the measured flux. If the flux from a source at redshift $z$ is attenuated by a factor $P(z)$ the observed luminosity distance {\it increases} as
\begin{equation}
d_L^{\,\rm obs}(z)  = \frac{d_L(z)}{\sqrt{P(z)}}\,.
\end{equation}

The apparent extension of the luminosity distance by photon interactions and oscillations has been investigated in the context of axion-like particles~\cite{Csaki:2001yk,Deffayet:2001pc,Mirizzi:2006zy}, hidden photons~\cite{Evslin:2005hi}, and also chameleons~\cite{Burrage:2007ew}. One of the main attractions of these models is the possibility that the conclusions about the energy content of our Universe drawn from the Hubble diagram can be dramatically altered. In particular, a contribution of dark energy as the source of the observed accelerated late-time expansion of the Universe could be completely avoided by a strong dimming effect.

In this paper we will consider the possibility to constrain mini-charged particles (MCPs) by their effect on the luminosity--redshift relation in the standard cosmological model. At first glance, these hypothetical particles seem to be at odds with the observation that all known elementary particles obey the principle of charge quantization, {\it i.e.} all charge ratios appear to be rational numbers close to unity. Moreover, there are attractive extensions of the Standard Model, in particular grand unified theories that enforce this quantization naturally. However, charge quantization need not to be a fundamental principle and it is possible that extensions of the Standard Model include very light particles with extremely small electromagnetic charges. In particular, MCPs may arise naturally in extensions of the Standard Model via gauge kinetic mixing~\cite{Holdom:1985ag} or in extra-dimensional scenarios~\cite{Batell:2005wa}. Typical predicted values for the mini-charge $\epsilon$ cover a wide range between $10^{-16}$ and $10^{-2}$ in terms of the electron's charge~\cite{Holdom:1985ag,Batell:2005wa,stringref}. 

We will focus in this paper on SN dimming in a simple extension of the Standard Model by one additional MCP, either a Dirac spinor or a scalar. We will show that, even with the rather large observational errors involved in redshift surveys, the limit on the charge $\epsilon$ of very light MCPs is about two order of magnitude stronger than laboratory bounds~\cite{Ahlers:2007qf}. This supplements comparable (and even stronger) bounds from the study of cosmological and astrophysical environments (for reviews see~\cite{Davidson:2000hf}). Note, that non-minimal MCP models, in particular kinetic mixing scenarios with additional abelian gauge bosons can partially alter these charge bounds. We will also comment on this effect on our limits from SN dimming. Finally, we briefly speculate about a strong dimming effect as an alternative to dark energy.

\section{SN Dimming by MCPs}

Pair production of MCPs by star light may take place via interactions in the cosmic microwave background (CMB) or via photon decay in the inter-galactic (IG) electron plasma and magnetic field. The latter process dominates in the high-frequency ($m_\epsilon\ll\omega$) and strong-field ($m_\epsilon^2\ll\epsilon e {\rm B_{IG}}$) limit with an average IG magnetic field strength\footnote{To be more precise, ${\rm B_{IG}}$ denotes the IG field component perpendicular to the line of sight.} ${\rm B_{IG}}$ of the order of $1$nG~\cite{Kronberg:1993vk}. The MCP pair production rate for unpolarized light is given as\footnote{We work in natural Heaviside-Lorentz units with $\hbar=c=1$, $\varepsilon_0=\mu_0=1$, $\alpha=e^2/(4\pi)\simeq1/137$ and $1~{\rm G}\simeq1.95\times10^{-2}{\rm eV}^2$.}  
\begin{align}\label{GIG}
\Gamma_{\rm B} = \sqrt{\pi}\alpha^{3/2}(1+z)^\beta\frac{\epsilon^3\,{\rm B_{IG}}}{m_\epsilon}\,\langle T\rangle\,,
\end{align}
where we assume ${\rm B_{IG}}(z) = (1+z)^\beta\,{\rm B_{IG}}$ with $\beta\simeq0$ for `replenishing' and $\beta\simeq2$ for `adiabatic' magnetic field expansion.
The polarization-averaged quantity $\langle T\rangle = (T_\parallel+T_\perp)/2$ (see Refs.~\cite{Gies:2006ca}) can be parametrized by the dimensionless parameter
\begin{align}\label{lambda}
\chi &\equiv 3\sqrt{\pi\alpha}(1+z)^{1+\beta}\,\frac{\epsilon\,\omega\,{\rm B_{IG}}}{m_\epsilon^3}\\\nonumber
&\simeq8.86\,(1+z)^{1+\beta}\,\frac{\epsilon_{-6}\,\omega_{\rm eV}\,{\rm B}_{\rm IG, nG}}{m^3_{\epsilon,\mu{\rm eV}}}\,,
\end{align}
where we have introduced the abbreviations $\epsilon = \epsilon_{n}10^{n}$, $\omega=\omega_{\rm eV}\,{\rm eV}$, {\it etc.} Asymptotically, $\langle T\rangle$ is given by
\begin{align}
\langle T\rangle = \begin{cases}a_-\frac{3}{8}\sqrt{\frac{3}{2}}\exp\left(-\frac{4}{\chi}\right)\,&\text{for}\,\,\,\chi\ll1\,,\\
a_+\frac{5}{6}\frac{2\pi}{\Gamma\left(\frac{1}{6}\right)\Gamma\left(\frac{13}{6}\right)}\chi^{-1/3}\,&\text{for}\,\,\,\chi\gg1\,,
\end{cases}
\end{align}
with $a_\mp=1$ for Dirac spinors and $a_\mp=(1/6,1/5)$ for scalars.
For $\chi\gg1$ -- corresponding to the high-frequency and strong-field limit -- the pair production rate~(\ref{GIG}) is independent of the MCP mass and can be written
\begin{equation}\label{GIG2}
\Gamma_{\rm B}\simeq6.6\,{\rm Gpc}^{-1}\,\left(\frac{a_+^3\,(1+z)^{2\beta-1}\,\epsilon_{-8}^8\,{\rm B}^2_{\rm IG, nG}}{\omega_{\rm eV}}\right)^{1/3}\,.
\end{equation} 
Note, that in this limit the MCP pair production rate is stronger for lower photon frequencies, resulting in a \emph{blueing} of distant star light.

%%%%%%%%%%%%%%%%%%%%%%%%%%%%%%
\begin{figure}[t]
\begin{center}
\includegraphics[width = \linewidth, bb=151 498 503 680]{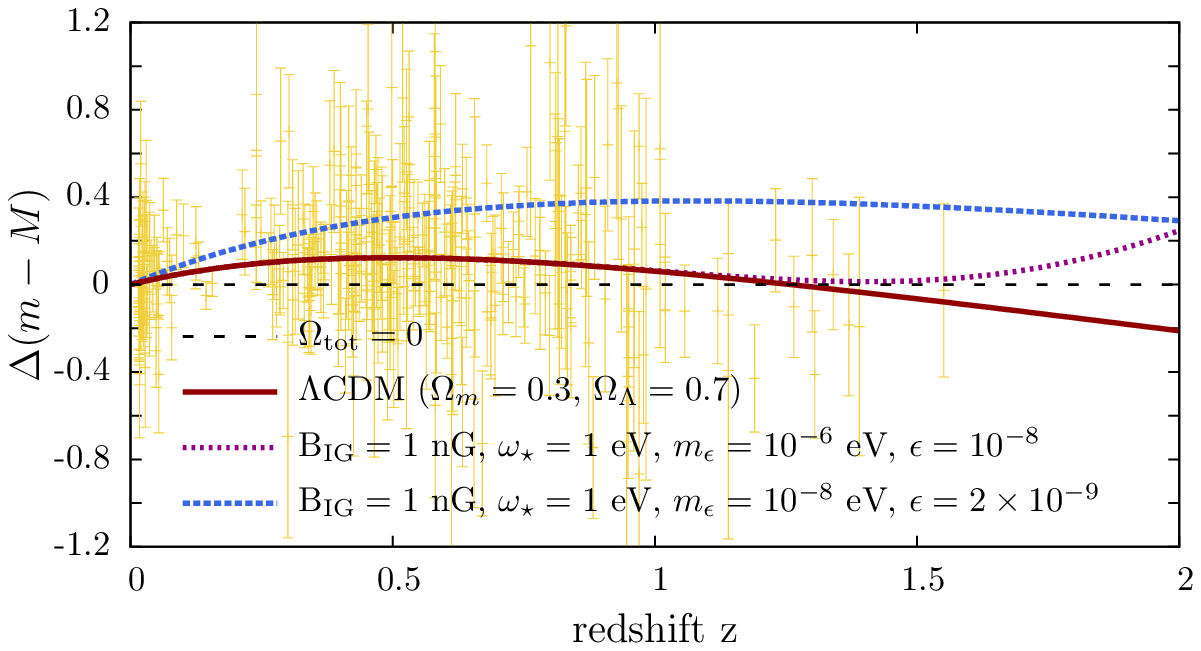}\\[0.3cm]
\includegraphics[width = \linewidth, bb=165 495 503 679]{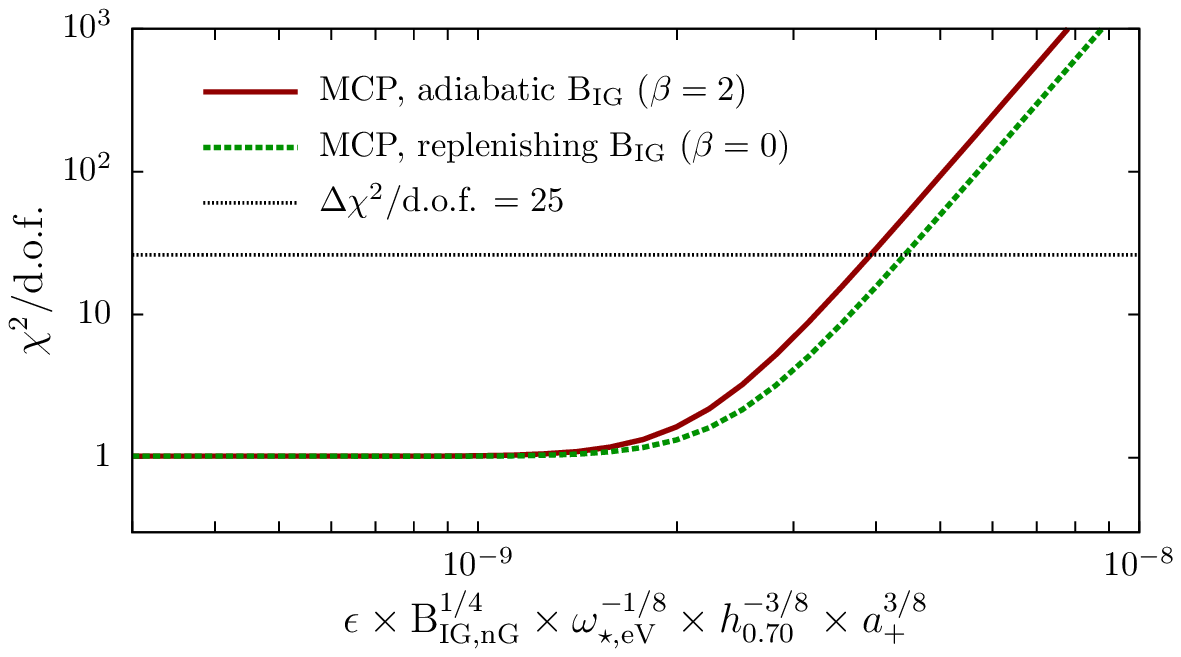}
\end{center}
\vspace{-0.3cm}
\caption[]{{\bf Upper Panel:} Hubble diagram showing the SNe Ia `union' compilation from Ref.~\cite{Kowalski:2008ez}. The luminosity distance $d_L$ is shown by the difference $\Delta(m-M)$ to an empty $\Omega_{\rm tot}=0$ flat universe. We show the effect of an MCP spinor with two different combinations of $m_\epsilon$ and $\epsilon$ on the luminosity distance of sources observed in a frequency interval centered at $\omega_\star$. {\bf Lower Panel:} The reduced $\chi^2$ of the SNe Ia `union' compilation~\cite{Kowalski:2008ez} with MCP production in the limit $m_\epsilon\to0$ assuming a `replenishing' ($\beta=0$) and an `adiabatic' ($\beta=2$) IG magnetic field. We show the deviation \mbox{$\Delta\chi^2$/d.o.f.~$=25$} relative to the $\Lambda$CDM model indicating the strength of a \mbox{$5\sigma$-deviation}. The MCP model is parametrized by the combination \mbox{$\epsilon\times{\rm B_{IG, nG}^{1/4}}\times\omega_{\star,\rm eV}^{-1/8}\times h_{0.70}^{-3/8}\times a_+^{3/8}$} in the massless limit.}\label{dimming}
\end{figure}
%%%%%%%%%%%%%%%%%%%%%%%%%%%%%%

The differential flux of photons from a source at redshift $z$ is reduced by the exponential factor
\begin{equation}
P(z) = \exp\left(-\int_0^z\D\ell\,\Gamma_{\rm B}(\omega)\right)\,,
\end{equation}
where the propagation distance $\ell$ is given by \mbox{$\D\ell = H(z)(1+z)\D z$} with Hubble parameter $H$. Hence, the modified luminosity distance~(\ref{dL}) of a source observed in a (small) frequency band centered at $\omega_\star$ increases as
\begin{equation}\label{approx}
d_L^{\,\rm obs}(z) \simeq {d_L(z)}\, \exp\left(\frac{1}{2}\int_0^z\frac{\D z'\,\Gamma_{\rm B}(z',\omega_\star)}{H(z')(1+z')}\right)\,,
\end{equation} 
where in a homogeneous and isotropic universe the luminosity distance is predicted as
\begin{equation}
d_L(z)  = (1+z)\,a_0\,\Phi\left( \int_0^z\frac{\D z'}{a_0H(z')}\right)\,,
\end{equation}
with $a_0^{-1} = H_0\sqrt{|1-\Omega_{\rm tot}|}$ and $\Phi_k(\xi)=(\sinh\xi,\xi,\sin\xi)$ for spatial curvature \mbox{$k=-1,0,1$}, respectively.

As an example, the upper panel of Fig.~\ref{dimming} shows the contribution from two MCP set-ups in the $\Lambda$CDM model with $\Omega_m\simeq0.3$ and $\Omega_\Lambda\simeq0.7$. The Hubble parameter at redshift $z$ is given by \mbox{$H^2(z) = H_0^2(\Omega_m(1+z)^3+\Omega_\Lambda)$} where the present Hubble expansion is \mbox{$H_0 = h\,100\,{\rm km}\,{\rm s}^{-1}\,{\rm Mpc}^{-1}$} with $h\simeq0.7$. The luminosity distance is shown as the difference between the measured apparent magnitude $m$ and the known absolute magnitude $M$ given by 
\begin{equation}
m-M = 5\log_{10}d_{L,{\rm Mpc}}+25\,.
\end{equation}

The MCP pair production rate~(\ref{GIG}) only depends on the MCP charge $\epsilon$ in the limit $\chi\ll1$. In the lower panel of Fig.~\ref{dimming} we show the modified reduced $\chi^2$ value in comparison with the SNe Ia `union' compilation~\cite{Kowalski:2008ez} for a varying MCP charge $\epsilon$ in the limit of small MCP masses. From this we can derive an upper bound on the charge of MCPs with mass $m_\epsilon\lesssim10^{-7}$~eV ({\it cf.}~Eq.~(\ref{lambda})) of\,\footnote{Note, that for scalar MCPs the bound is weaker by a factor $a_+^{-3/8}\simeq2$ compared to the case of Dirac spinors.}
\begin{equation}\label{limit}
\epsilon\lesssim4\times10^{-9}\times{\rm B_{IG, nG}^{-1/4}}\times\omega_{\star,\rm eV}^{1/8}\times h_{0.70}^{3/8}\times a_+^{-3/8}\,.
\end{equation}
For larger MCP masses \mbox{$m_\epsilon\gtrsim10^{-7}$~eV} the rate $\Gamma_{\rm B}$ becomes mass-dependent and the MCP dimming effect sets in at higher redshift ({\it cf.}~upper panel of Fig.~\ref{dimming}). In this mass-region the sensitivity of the Hubble diagram to MCP production is limited by the observational errors. 

The limit (\ref{limit}) improves laboratory bounds on MCP charges by about two order of magnitude~\cite{Ahlers:2007qf} and supplements other cosmological and astrophysical bounds in the range $10^{-7}\lesssim\epsilon\lesssim10^{-14}$ coming from the effect of MCPs on big bang nucleosynthesis or on the evolution of stellar objects like SN 1987A, red giants and white dwarves (for reviews see Ref.~\cite{Davidson:2000hf}). However, it has been argued that these bounds could be (partially) avoided in non-minimal hidden sector models~\cite{Jain:2005nh}. 

There are also strong bounds $\epsilon\lesssim10^{-8}$ from the study of the cosmic microwave background~\cite{Melchiorri:2007sq}, which can even be extended to $\epsilon\lesssim10^{-9}$ in kinetic mixing scenarios considering the scattering processes involving the additional hidden photons. The effect of the hidden photon in SN dimming is the exact \emph{opposite}. A photon emitted from the source is only initially in its electromagnetic interaction eigenstate.  After a distance of the order of $\epsilon^2/\Gamma_{\rm B}$, the state has evolved into a superposition of the photon and hidden photon states, whose combined coupling to the MCP is drastically reduced\footnote{The coupling vanishes up to contributions proportional to the plasma frequency of the IG electron plasma.}~\cite{Ahlers:2007rd}. Hence, a kinetic MCP production rate $\Gamma_{\rm B}$ as low as the Hubble expansion rate is not observable in this scenario and our bound does not apply in this case.

%%%%%%%%%%%%%%%%%%%%%%%%%%%%%%
\begin{figure}[t]
\begin{center}
\includegraphics[width = \linewidth, bb=151 498 503 680]{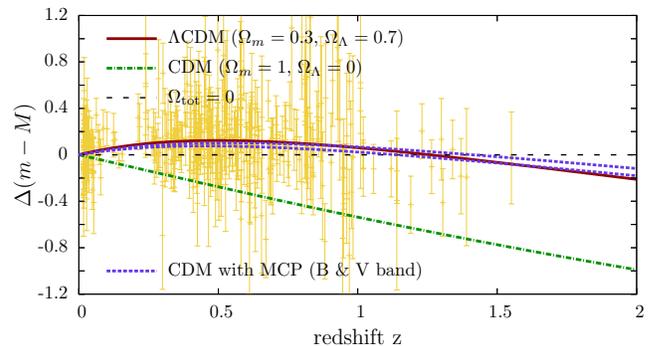}
\end{center}
\vspace{-0.3cm}
\caption[]{As upper panel of Fig.~\ref{dimming}, but now showing also a flat CDM model with $\Omega_m=1$ and $\Omega_\Lambda=0$. We consider a mini-charged Dirac spinor with charge $\epsilon=3.1\times10^{-9}$ and mass $m_\epsilon\ll10^{-7}$~eV and show the dimming for the B ($\lambda_\star\simeq440$~nm, lower line) and V ($\lambda_\star\simeq550$~nm, upper line) band. We also assume an adiabatically expanding ($\beta=2$) IG magnetic field with strength B=$1$~nG. The (absolute) relative difference of $\Delta(m-M)$ between the B and V band is $\lesssim0.06$ for $z\lesssim1.8$.}\label{DEexample}
\end{figure}
%%%%%%%%%%%%%%%%%%%%%%%%%%%%%%

\section{Strong dimming as an alternative to dark energy?}

We have shown that light MCPs with a charge larger than a few$\times10^{-9}$ can have an observable effect on the luminosity distance measured by SN surveys. So far we have only considered the limits on possible values of MCP charges and masses that arise from a comparison with the $\Lambda$CDM model. However, one might also ask if the dimming effect of MCPs could be significant enough to change the usual conclusion of the underlying cosmological model. In particular, the accelerated late-time expansion of the Universe observed by SN surveys could be attributed to a strong MCP dimming in a flat CDM model with $\Omega_m\simeq1$ and $\Omega_\Lambda\simeq0$.

Before we start to sketch a possible model, we would like to stress that the $\Lambda$CDM model is also (indirectly) substantiated by other cosmological observations, in particular by the analysis of angular anisotropies in the CMB and of spatial correlations in the large-scale distribution of galaxies~\cite{Amsler:2008zzb}. However, these observations can be fitted equally satisfactorily in alternative models which have a small component of neutrino hot dark matter and invoke non scale-free primordial density fluctuations (for a critical review see Ref.~\cite{Sarkar:2007cx}).  

Figure~\ref{DEexample} shows an example of a mini-charged Dirac spinor with charge $\chi=3.1\times10^{-9}$ and mass $m_\epsilon\ll10^{-7}$~eV in a CDM model. We assume an adiabatically expanding ($\beta=2$) IG magnetic field with strength ${\rm B_{IG}}=1$~nG. The apparent luminosity distance (\ref{approx}) is practically indistinguishable from the $\Lambda$CDM prediction. Note, however, that the photon absorption in the background magnetic field is achromatic since $\Gamma_{\rm B}\sim\omega^{-1/3}$ in the $\chi\gg 1$ region. This produces a \emph{negative} colour-excess between the B and V band of the form,
\begin{align}
E[{\rm B-V}] &\equiv \Delta(m-M)_{\rm B} - \Delta(m-M)_{\rm V}\\\nonumber
&\simeq-0.15\left(1-(1+z)^{-1/2}\right)\,,
\end{align}
which is also indicated in Fig.~\ref{DEexample}. For $z\gtrsim0.6$ the colour-excess is $|E[{\rm B-V}]|\gtrsim0.03$, which seems already to challenge the observed colour-excess of high-redshift SNe~\cite{Riess:1998cb,Kowalski:2008ez} (see also discussions in Refs.~\cite{Deffayet:2001pc,Mirizzi:2006zy}). Though this might constrain SN dimming by MCPs as a (full) dark energy alternative, it should be possible to derive even stronger bounds on MCP charges by this achromaticity.

\section{Conclusions}

The luminosity--redshift relation of cosmological standard candles provides information about the energy composition and geometry of our Universe. The long distance covered by photons from these sources is sensitive to the production of hypothetical weakly interacting and light particles in the inter-galactic environment. We have shown that mini-charged Dirac spinors with mass $m_\epsilon\lesssim10^{-7}$~eV and charge $\epsilon\gtrsim4\times10^{-9}$ are excluded by their dimming of SNe in conflict with the \mbox{luminosity--redshift} relation in the cosmological `concordance model'. This bound supplements other strong limits on MCP charges from cosmological and astrophysical environments.

On the other hand, if the dimming by MCP pair production is strong, the cosmological interpretation of SN surveys could be considerably modified. We have sketched a MCP model with an adiabatically expanding inter-galactic magnetic field of $1$~nG that reproduces the observed luminosity distance of SNe from a CDM model with $\Omega_m=1$ and $\Omega_\Lambda=0$. A characteristic feature of this SN dimming mechanism is a \emph{blueing} of the star light, giving a \emph{negative} colour-excess with $|E[B-V]|\lesssim0.06$ for redshift $z\lesssim1.8$.
 
\section*{Acknowledgements}
\vspace*{-0.3cm}
\noindent I would like to thank Arman Shafieloo for discussions, Subir Sarkar for helpful comments on an earlier version of the manuscript and Joerg Jaeckel, Javier Redondo and Andreas Ringwald for the nice collaboration and useful comments on the manuscript, in particular for pointing out the reduced dimming effect in gauge kinetic mixing scenarios. I acknowledge support by STFC UK (PP/D00036X/1).

\end{document}